\documentclass[aps,prb,twocolumn,groupedaddress]{revtex4-2}
\usepackage[english]{babel}
\usepackage{bm}
\usepackage{amsmath}
\usepackage[dvips]{graphicx}
\usepackage[colorlinks=true, allcolors=blue]{hyperref}
\usepackage{diagbox}

\begin{document}
\title{Nonlinear Schr\"odinger equation for a two-dimensional plasma: the analysis of solitons, breathers, and plane wave stability}

\author{A.~A.~Zabolotnykh}
\email{zabolotnykh@phystech.edu}
\affiliation{Kotelnikov Institute of Radio-engineering and Electronics of the RAS, Mokhovaya 11-7, Moscow 125009, Russia}

\begin{abstract}
    We analytically study nonlinear quasi-monochromatic plasma waves in a two-dimensional (2D) electron system (ES) set between the two metal electrodes (gates). We derive a nonlinear Schr\"odinger equation for a slow-varying envelope to describe the waves. We find it to be of either focusing or defocusing type depending on the parameter $qd$, where $q$ is the carrier wave vector and $d$ is the distance between the 2DES and the gates. When $qd<1.61$, we have the defocusing-type equation with the solutions in the form of dark plasma solitons appearing against the background of the stable plane waves. Conversely, for $qd>1.61$, the focusing-type equation has the solutions in the form of bright solitons, and the plane waves are unstable.  We also address the appearance of the simplest type of breathers in the latter case. A detailed description of the resultant nonlinear waves is given based on the parameters of the two-dimensional electron system.
\end{abstract}
\maketitle
\section{Introduction}
Studies of charge density excitations (or plasma waves) in low-dimensional systems have been carried on for over half a century \cite{Stern1967,Chaplik1972,Grimes1976,Allen1977,Theis1977}, and they still attract great interest. For the most part, it is because the 2D plasma waves can be applied in detecting THz and sub-THz radiation~\cite{Dyakonov1993,Satou2003, Shaner2005,Dyakonov2005, Knap2009,Muravev2012,Lusakowski2016, Bandurin2018,kapralov2020}. Along with the applied studies, there has been carried out intriguing fundamental research, for example, as reported in recent papers on plasmon-assisted compression of light~\cite{Iranzo2018,Koppens2020}, relativistic plasma waves \cite{muravev2015novel,Andreev2021,Andreev2021magnetodispersion}, nonlocal 2D plasma excitations~\cite{bandurin2022,kapralov2022}, the realization of the hydrodynamic regime~\cite{Gurzhi1968,Jong1995,Gurzhi1995}, etc.

Although nonlinear effects in 2DESs, for instance, the generation of a rectified current during the propagation of monochromatic waves, have been given a rather detailed consideration, see, for example~\cite{Knap2009}, as a rule, they are examined within the framework of a conventional perturbation theory and/or through an iterative process. However, to the best of our knowledge, the dynamics of the nonlinear waves themselves have not been studied.

Thus, in this paper, we propose to look at the nonlinear dynamics of quasi-monochromatic plasma waves from a more basic perspective. The standard method of describing the evolution of a weakly nonlinear quasi-monochromatic wave relies on the nonlinear Schr\"odinger equation (NLSE) formulated for the slow-varying envelope of the wave. The NLSE can describe a number of different nonlinear waves, like those on the water surface~\cite{osborne2002}, in fiber optics~\cite{copie2020}, in Bose-Einstein condensates~\cite{frantzeskakis2010}, in modulated structures~\cite{Kartashov2011}, etc. Nevertheless, as far as we know, the NLSE for the plasma waves in a 2DES so far has not been derived and analyzed.

It should be noted, that although the conventional iterative method describes sufficiently well the generation of high harmonics and the rectified response of the system, this approach does not capture slow variations in the amplitudes of plasma waves (compared to the period of the wave, $2\pi/\omega$). Consequently, it cannot adequately describe the evolution of nonlinear waves on large time scales. That is why, more accurate approach, called the multiple scales method, which enables the characterization of slow changes in the amplitude of the waves propagating in a system, is employed in the reported study to derive the NLSE and describe the dynamics of the envelope of the quasi-monochromatic plasma waves.

\begin{figure}[t]
\includegraphics[width=1.0\columnwidth]{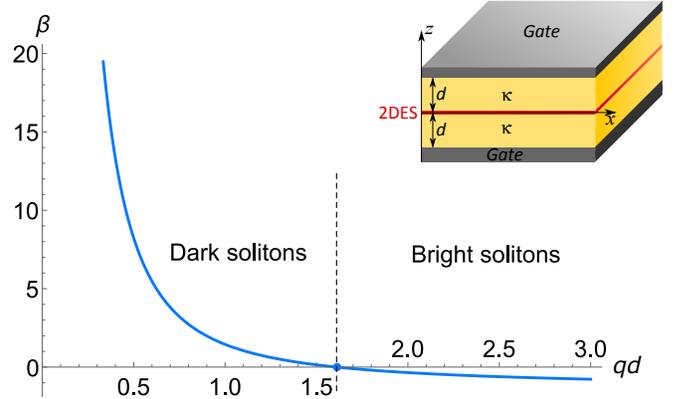}
    \caption{\label{fig:nonlinear} Dependence of the nonlinear-term coefficient, $\beta$, in (\ref{NLSE}), which is defined by Eq.~(\ref{bet}), on the dimensionless parameter $qd$, where $q$ is the carrier wave vector and $d$ is the distance between the 2DES and the gate. The coefficient vanishes at $qd\approx 1.61$. As $qd$ grows infinitely large, $\beta$ tends to the value of $-3/2$. The inset shows the schematic of the 2DES setup under consideration.
    }
\end{figure}

In the present work, we deduce and explore the NLSE for the plasma waves in a 2DES sandwiched between the two gates, as depicted in the inset of Fig.~\ref{fig:nonlinear}. In regard to relating the investigated scheme depicted in the inset to realistic systems, it should be noted that a heavily doped substrate (separated from the 2D system by the dielectric with the thickness $d$) can act as one of the gates. Therefore, real systems would not require creating two actual metal gates. It would suffice to fabricate only one, while a heavily doped substrate could play the role of the other. 

It should be noted that throughout the paper, by 2DESs we refer to the semiconductor heterostructures and quantum wells based primarily on GaAs/AlGaAs compounds. However, in principle, the phenomena reported in the paper can also be achieved in conducting 2D materials such as graphene, for additional discussion, please see the end of Sec.~\ref{EST}.

We find that the NLSE for 2D plasma waves can be of focusing or defocusing type, as determined by the parameter $qd$, with $q$ being the carrier wave vector and $d$ being the distance between the 2DES and the gates. In particular, $qd>1.61$ leads to the focusing NLSE with solutions in the form of bright solitons, in which case the plasma plane waves are unstable. Otherwise, $qd<1.61$ results in the defocusing NLSE with the dark plasma solitons as solutions, arising against the background of the stable plane waves.

Curiously, the NLSE for the water-surface waves has a similar behavior, with a comparable threshold value of the analogous parameter $qh=1.363$~\cite{hasimoto1972}, where $q$ is the carrier wave vector and $h$ is the water depth. In that case, it defines the borderline between the shallow-water ($qh<1.363$) and deep-water ($qh>1.363$) conditions.

\section{Nonlinear Schr\"odinger equation for the plasma waves in a 2DES}

Consider a 2D system placed between the two ideal metal gates positioned above and below it, as illustrated in the inset of Fig.~\ref{fig:nonlinear}. Hence, we look for the waves that propagate along the $x$-axis and are uniform along the $y$-axis. For the analytical description of the plasma waves in the 2DES, we follow a standard hydrodynamic approach~\cite{Dyakonov1993,Govorov1999}, which applicability is examined in Section~\ref{HDA}. The approach is based on using the Euler equation for the velocity of electron flow $v(x,t)$ and the continuity equation for the deviation in the electron concentration $n(x,t)$ from its equilibrium value $n_0$ along with the Poisson equation for the potential $\varphi(x,t)$ taken at the 2DES plane $z=0$:
\begin{equation}\label{Eq:basic}
\begin{split}
     &\partial_t v(x,t) +\partial_x \frac{v^2(x,t)}{2}=\frac{e}{m}\partial_x\varphi(x,t), \\
     &\partial_t n(x,t) +\partial_x \left[ (n_0+n(x,t))v(x,t)\right]=0,\\
     &\varphi(x,t)=\frac{-e}{\kappa}\int^{+\infty}_{-\infty}G(x-x')n(x',t)dx',
\end{split}
\end{equation}
where $\kappa$ is the dielectric permittivity of the medium between the 2DES and the gates, $e$ is the elementary charge, and $m$ is the effective mass of electrons. Here, the Poisson equation is given in the integral form in terms of the Green's function, $G(x)$, of the Laplace operator, with the boundary conditions of vanishing potential at the gate surfaces $z=\pm d$ in the $xy$-plane. The Fourier transform of the Green's function, $G(q)$, can be written, as:
\begin{equation}
\label{Eq:Gf}
     G(q)=\frac{2\pi}{q} \tanh qd,
\end{equation}
the derivation of $G(q)$ is given in Appendix~\ref{App:Green}.
Since the spectrum of the linear plasma waves in a 2DES, $\omega(q)$, is related to $G(q)$ through $\omega^2(q)=e^2 n_0q^2G(q)/m\kappa$ (see Appendix~\ref{App:Green}), we arrive at the following expression for the spectrum~\cite{Dahl1977,Ando1982}:
\begin{equation}
    \label{Eq:spectrum}
    \omega^2(q)=\frac{2\pi e^2 n_0}{m\kappa}q \tanh{qd},
\end{equation}
where $q$ is the modulus of the wave vector directed, in this case, along the $x$-axis.

To find the NLSE for the plasma waves, we solve the equations in~(\ref{Eq:basic}) employing the multiple scales method~\cite{hasimoto1972,davey1974}. Within that framework, the functions describing the wave are expanded into a series governed by the small parameter $\varepsilon$: 
\begin{equation}
\label{variables}
\begin{split}
 &v(x,t)=\sum\limits^{\infty}_{k=1} \varepsilon^k v_k(x_0,x_1,x_2,...,t_0,t_1,t_2,...),\\
 &n(x,t)=\sum\limits^{\infty}_{k=1} \varepsilon^k n_k(x_0,x_1,x_2,...,t_0,t_1,t_2,...),\\
 &\varphi(x,t)=\sum\limits^{\infty}_{k=1} \varepsilon^k \varphi_k(x_0,x_1,x_2,...,t_0,t_1,t_2,...).
 \end{split}
\end{equation}
Here, the parameter $\varepsilon$ is used to separate different orders of nonlinearity. Namely, the linear response is defined by $v_1$, $\varphi_1$, and $n_1$, whereas the quadratic response is defined by $v_2$, $\varphi_2$, and $n_2$ that contain the rectified contribution and the contribution of the second harmonic $\propto 2\omega$, etc. However, in addition to the rectification and generation of high-order harmonics, the nonlinearities in Eqs.~(\ref{Eq:basic}) lead to slow (in the sense that nonlinearity is relatively small) variation in the amplitudes of the first and other harmonics. This slow variation is described in the framework of the multiple scales method through the introduction of (generally speaking, phenomenological) additional arguments $x_k$ and $t_k$ for $k=1,2,..$. Thereby, different arguments in functions $v_k$, $\varphi_k$, and $n_k$ are used to distinguish between the fast and slow dynamics of the quasi-monochromatic wave under consideration. Namely, the arguments $x_0$ and $t_0$ describe the fast carrier of the wave and have the same order as the ''real'' $x$ and $t$. On the other hand, the arguments, $x_k$ and $t_k$, with $k\ge 1$, are qualitatively of the order of $\varepsilon^k x$ and $\varepsilon^k t$, describing the slow evolution of the envelope.
In the given framework, the differential operators become as follows:
\begin{equation}\label{Eq:Dif}
\begin{split}
 &\partial_x \to \partial_{x_0}+\varepsilon \partial_{x_1}+\varepsilon^2 \partial_{x_2} +...\\
 &\partial_t \to \partial_{t_0}+\varepsilon \partial_{t_1}+\varepsilon^2 \partial_{t_2} +...
\end{split}
\end{equation}

Applying the multiple scales method to the first two equations in (\ref{Eq:basic}) using Eqs.~(\ref{variables}) and (\ref{Eq:Dif}) is fairly straightforward. However, as for the Poisson equation, the application of the method needs more detailed consideration, which is given in Appendix~\ref{App:ML}.

The essential idea of the multiple scales method is successively setting to zero the coefficients of the powers of the small parameter $\varepsilon$, i.e. $\varepsilon$, $\varepsilon^2$, $\varepsilon^3$, and so on, which leads to a series of equations for $v_k$, $n_k$, and $\varphi_k$. Hence, by substituting Eqs.~(\ref{variables}), (\ref{Dif_ML}), and (\ref{Dif_ops}) into the first two expressions in (\ref{Eq:basic}) and Eq.~(\ref{Eq:Psplit}), respectively, we obtain the desired set of equations for $v_k$, $n_k$, and $\varphi_k$. A detailed application of the method is given in the Appendix~\ref{App:EE}, while below only the main relations are given.

Retaining only terms proportional to $\varepsilon$, which corresponds to the case of linear waves, 
and looking for the concentration deviation $n_1(x_0,x_1,...,t_0,t_1,...)$ of the following form:
\begin{equation}
\label{defin_n}
    n_1=A(x_1,x_2,...,t_1,t_2,...) e^{iqx_0-i\omega t_0}+c.c.,
\end{equation}
where $A$ is a 'slow varying' complex amplitude, as it does not depend on the ‘fast’ variables $x_0$ and $t_0$, we arrive at:
\begin{equation}
\label{eps1}
    \left(\omega^2 - v_p^2 q^2 G_0(q)\right)A \exp(i\theta)=0,
\end{equation}
where $v_p=\sqrt{2 \pi e^2 n_0 d /m\kappa}$ is the velocity of the linear plasma waves (\ref{Eq:spectrum}) in the long-wavelength limit $qd \ll 1$, $\theta=q x_0-\omega t_0$, and $G_0(q)=\tanh(qd)/qd$. To satisfy Eq.~(\ref{eps1}), the expression in brackets must be equal to zero. Therefore, as expected, Eq.~(\ref{eps1}) relates $\omega$ and $q$, i.e. it leads to the dispersion relation (\ref{Eq:spectrum}) for linear plasmons in a 2DES.

Now, let us proceed to the equations that follow from the vanishing of the coefficient of $\varepsilon^2$, which are given in Appendix~\ref{App:EE}, see Eqs.~(\ref{Eq:e2}).
The key point here is that in the framework of the multiple scales method, all the corrections $n_2, n_3,...$ should be finite, without unbounded growth at $t_0 \to \pm \infty$ or $x_0 \to \pm \infty$, that is why no 'resonant' (secular) terms $\propto \exp(i\theta)$ are expected in the right-hand side of the equation  for corrections $n_2, n_3,..$. 

Hence, setting to zero the coefficient of the term $\propto \exp(i\theta)$ in the right-hand side of the equation for $n_2$, see Eq.~(\ref{Eqn2}) in Appendix~\ref{App:EE}, results in the following equation for the envelope $A$:
\begin{equation}
\label{envelope_vel}
    \partial_{t_1} A +v_{gr} \partial_{x_1} A=0,
\end{equation}
where $v_{gr} =\partial \omega/\partial q$ is the group velocity of plasma waves. The obtained equation clearly indicates that the envelope propagates with group velocity, as expected.

Last, we derive the equations that follow from the vanishing of the coefficient of $\varepsilon^3$ in Eqs.~(\ref{Eq:basic}) and (\ref{Eq:Psplit}) after the substitution of (\ref{variables}), (\ref{Dif_ML}) and (\ref{Dif_ops}), which are included as Eqs.~(\ref{App:e3}) in Appendix~\ref{App:EE}. 
Setting to zero the coefficient of the secular term $\propto \exp(i\theta)$ in the right-hand side of the equation for the correction $n_3$ (\ref{App:Eqn3}) yields the NLSE of the following form:
\begin{equation}
\label{NLSE}
    i \left(\partial_{t_2} + v_{gr}\partial_{x_2} \right)A +
  \frac{\omega_{qq}(q)}{2}
    \partial_{x_1}^2 A +\frac{\omega(q)}{n_0^2}\beta(q) |A|^2 A =0,
\end{equation}
where $\omega_{qq}(q)=\partial^2 \omega/\partial q^2$, while the coefficient of the nonlinear term, $\beta(q)$, is defined as:
\begin{equation}
\label{bet}
    \beta(q) = \frac{4v_p^2+4v_{gr}\omega/q+\omega^2/q^2}{2(v_p^2-v_{gr}^2)} -\frac{5G_0(q)/2+2G_0(2q)}{2(G_0(q)-G_0(2q))}.
\end{equation}
Note that $\omega(q)$ is the dispersion law of plasmons in the 2DES~(\ref{Eq:spectrum}), $G_0(q)$ corresponds to the Fourier transform of Coulomb's law taken at 2DES plane (see Appendix~\ref{App:Green}), $v_p$ is the plasmon velocity in the long-wavelength limit $qd \ll 1$, $G_0(q)$ and $v_p$ are defined after Eq.~(\ref{eps1}), $v_{gr} =\partial \omega/\partial q$ is the group velocity of plasma waves, and $\omega_{qq}(q)=\partial v_{gr}/\partial q$ describes the change in the group velocity with the wave vector $q$. Also, according to (\ref{defin_n}), the amplitude of the concentration deviation, $n_1$, equals $2|A|$. 

Finally, Eq.~(\ref{NLSE}) can be rewritten in a more conventional form. We make the transition to the frame of reference moving with the group velocity, $v_{gr}$, and introduce the variables $T=t_2$ and $X=x_1-v_{gr}t_1$ (as well as $X_2=x_2-v_{gr}t_2$). As a result, in addition to the fact that Eq.~(\ref{envelope_vel}) is fulfilled, the NLSE appears as:
\begin{equation}
\label{NLSEf}
    i \partial_{T} A +\frac{\omega_{qq}(q)}{2}\partial_{X}^2 A +\frac{\omega(q)}{n_0^2}\beta(q) |A|^2 A =0.
\end{equation}

The dependence of the coefficient $\beta(q)$ on the parameter $qd$ is plotted in Fig.~\ref{fig:nonlinear}, clearly indicating the critical point $qd\approx 1.61$ where $\beta$ changes its sign, which is crucial for the type of the NLSE solutions.

On the whole, the resultant NLSE (\ref{NLSEf}) describes a variety of nonlinear quasi-monochromatic waves in a 2DES. In the present study, we focus in particular on a few basic phenomena that we find most interesting, namely, the plane wave stability, the 'dark' and 'bright' solitons, and the simplest type of breathers. These issues are considered in the next Section.

\section{NLSE solutions}
\subsection{The 'shallow' case of $qd<1.61$}
It has been shown that the stability of plane waves is determined by the sign of the product $\omega_{qq}\beta$~\cite{lighthill1965,bespalov1966,benjamin1967,zakharov1968}. Namely, a plane wave is stable provided that $\omega_{qq}\beta<0$, otherwise, when $\omega_{qq}\beta>0$, the plane wave is unstable due to the long-wavelength perturbations. In the case of the plasma waves in a 2DES, $\omega_{qq}$ is always negative, as follows from Eq.~(\ref{Eq:spectrum}). Therefore, the stability of plane waves depends solely on the coefficient $\beta$ in (\ref{bet}). As indicated in Fig.~\ref{fig:nonlinear}, we find $\beta$ to be positive for $qd<1.61$, which leads to stable plane waves.

\begin{figure}[t]
\includegraphics[width=1.0\columnwidth]{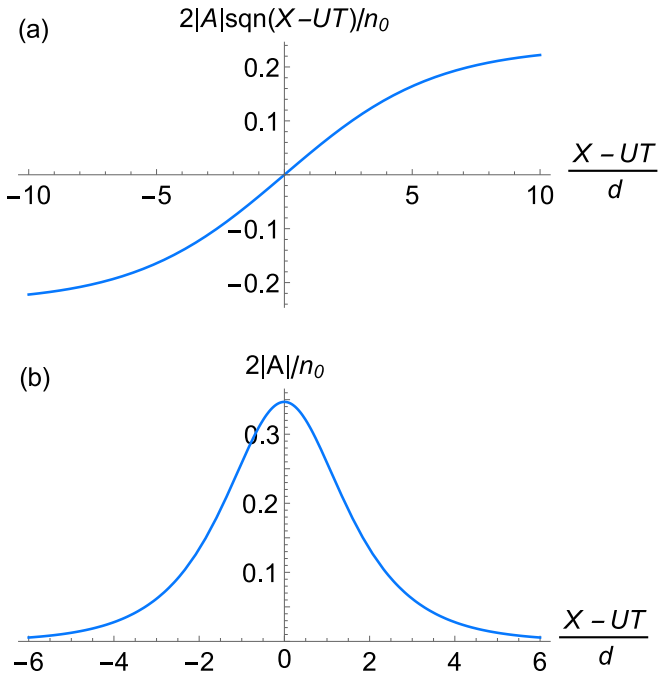}
    \caption{\label{fig:solitons} 
(a) The characteristic envelope $2|A|sqn(X-UT)/n_0$ for the dark soliton~(\ref{Eq:dark}), plotted for the dimensionless parameters $U/v_p=0.15$, $V/v_p=0.1$, and $qd=1.2$. (b) The characteristic envelope  $2|A|/n_0$ for the bright soliton~(\ref{Eq:bright}), plotted for $U/v_p=0.25$, $U/v_p=0.12$, and $qd=3$.
    }
\end{figure}

In that case, the so-called dark solitons can appear against the plane-wave background, manifested as the vanishing of the envelope function $A$. Since such solitons are relatively well established theoretically~\cite{zakharov1973,kivshar1998} and by experiment~\cite{emplit1987,krokel1988,weiner1988, shukla2006,heidemann2009,smirnov2006,frantzeskakis2010,Chabchoub2013}, only the major findings concerning them are reported in this paper. To begin with, we seek the solutions in the form of $A=a\exp(i\phi)$, where the amplitude $a=a(X-UT)$ and phase $\phi=\phi(X-VT)$ have the respective velocities $U$ and $V$. The fact that $U$ and $V$ are different distinguishes NLSE solitons from the conventional ones described, for instance, by Korteweg–De Vries equation~\cite{korteweg1895,Benjamin1972}. After the substitution of $a$ and $\phi$ into the NLSE (\ref{NLSEf}), we arrive at the following expression for the dark soliton~\cite{zakharov1973,kivshar1998}:
\begin{equation}
\label{Eq:dark}
    A=a_d \tanh\left(\frac{a_d}{n_0} \sqrt{\frac{\omega\beta}{|\omega_{qq}|}}(X-UT) \right)\exp\left(i\phi\right),
\end{equation}
where the amplitude $a_d$ and the phase $\phi$ are defined as:
\begin{equation}
    a_d=n_0\sqrt{\frac{U(2V-U)}{2|\omega_{qq}|\omega \beta}}\quad \text{and}\quad \phi =\frac{U(X-VT)}{|\omega_{qq}|}.
\end{equation}
Here, we note that dark solitons exist only when $2V>U$. The profile of the given soliton is shown in Fig.~\ref{fig:solitons}(a). It is also worth mentioning that the soliton under consideration is the simplest case in the whole family of gray solitons~\cite{zakharov1973,Chabchoub2013}. 

\subsection{The 'deep' case of $qd>1.61$}
When the coefficient $\beta$ is negative, the plane waves are unstable due to the long-wavelength perturbations, which is analogous to the Benjamin-Feir instability of the water-surface waves specified by the 'deep water' condition~\cite{lighthill1965,bespalov1966,benjamin1967,zakharov1968}. In this case of $qd>1.61$, the so-called bright solitons emerge. Unlike dark solitons, they can be formed, for example, as a result of the development of plane wave instability. 
As the solitons have been reviewed in detail in the literature~\cite{zakharov1971,Yuen1975,yuen1982}, here we include only the final expressions pertaining to our study. Hence, considering the solutions in the form of $A=a\exp(i \phi)$, with $a$ and $\phi$ introduced as in the previous subsection, leads to the following equation for the envelope function:
\begin{equation}
\label{Eq:bright}
    A=\frac{a_b\exp\left(iU(X-VT)/|\omega_{qq}|\right)}{ \cosh\left((X-UT)\sqrt{U(U-2V)}/|\omega_{qq}|\right)},
\end{equation}
where the amplitude $a_b$ is defined as:
\begin{equation}
    a_b= n_0 \sqrt{\frac{U(U-2V)}{\omega \beta \omega_{qq}}}.   
\end{equation}
In this case, it should be noted that bright solitons exist only if the condition $U>2V$ is satisfied. A typical profile of the bright soliton is shown in Fig.~\ref{fig:solitons}(b).

As the final step, we address the NLSE and some of its solutions in the limit $qd \to \infty$, i.e. for the ungated 2DES. After the introduction of the dimensionless variables:
\begin{equation}
   \overline{T} =-\frac{\omega T}{8}, \quad \overline{X}=qX, \quad \overline{A} = \sqrt{6} \frac{A}{n_0},
\end{equation}
the given NLSE~(\ref{NLSEf}) takes a standard form:
\begin{equation}
\label{NLSEd}
    i\partial_{\overline{T}}\overline{A}+ \partial^2_{\overline{X}} \overline{A} + 2|\overline{A}|^2\overline{A}=0.
\end{equation}
This equation has several interesting solutions including the ones like the Akhmediev breather~\cite{akhmediev1987} and the Kuznetsov-Ma breather~\cite{kuznetsov1977,ma1979}. In the present work, however, we consider a special type of solution called the Peregrine soliton \cite{peregrine1983}. Not only is it localized in space, as an ordinary soliton, but also in time, having just one pronounced maximum in its time-waveform. Hence, the solution to Eq.~(\ref{NLSEd}) can be written as:
\begin{equation}
\label{Peregrine}
    \overline{A}=\left(1-\frac{4+16 i \overline{T}^2}{1+4\overline{X}^2+16 \overline{T}^2}\right) \exp(2i\overline{T}).
\end{equation}
We note that along with the Peregrine soliton, there exist a whole series of space-time localized solutions~\cite{akhmediev2009,akhmediev1985,chabchoub2012}. Also, while the solution in~(\ref{Peregrine}) is given for the unity amplitude, it can be modified for the amplitude $A_0$ by making the substitutions $\overline{A} \to \overline{A}/A_0$, $\overline{X} \to \overline{X}A_0$, and $\overline{T} \to \overline{T}A_0^2$, which does not change the NLSE~(\ref{NLSEd}). 

It should be stressed that although the NLSE solutions may seem exotic, overall, they properly describe the waves in actual physical systems. For instance, the above mentioned Peregrine breathers have been observed experimentally in fiber optics~\cite{kibler2010,dudley2019}, on a water surface~\cite{chabchoub2011}, and in multicomponent plasma~\cite{bailung2011}. This gives grounds to believe that the results considered in this section are relevant for real 2DESs. 

\section{Applicability of the hydrodynamic approach} \label{HDA}

First of all, it should be mentioned that semiconductor heterostructures and quantum wells, primarily, based on GaAs/AlGaAs compounds, are implied by 2DESs in the following discussion.

Now let us consider the applicability of the the hydrodynamic approach, which is used to describe the electron dynamics, see the first equation in (\ref{Eq:basic}). This approach is applicable provided that the time of electron-electron collisions, $\tau_{ee}$, is less than the electron relaxation time due to collisions with impurities and phonons, $\tau_{ip}$. The typical values of $\tau_{ee}$ and $\tau_{ip}$ (as well as the kinematic viscosity, $\nu$, and the electron mobility, $\mu$) for GaAs/AlGaAs quantum wells are given in Table~\ref{Table}. In the table, $\tau_{ee}$ and $\nu$ are estimated for the case of the degenerate system as follows (see, for example, Refs.~\cite{Koshelev2017,gusev2020}): 
\begin{equation}
\label{2D_par}
  \tau_{ee} = \frac{\hbar E_f}{(k_B T)^2},  \quad  \nu = \frac{1}{4} v_f^2 \tau_{ee},
\end{equation}
where $E_f=mv_f^2/2$ and $v_f=\hbar\sqrt{2 \pi n_0 }/m$ denote the Fermi energy and velocity, the spin and valley degeneracy factors were taken equal two and one, $n_0$ is the equilibrium electron concentration, and $\nu$ is the kinematic viscosity. The estimates of $\tau_{ee}$ listed in the table are obtained for the effective mass of electrons in AlGaAs quantum wells  $m=0.066m_0$, with $m_0$ being the mass of a free electron, at a typical concentration $n_0=5\cdot 10^{11}$ cm$^{-2}$, leading to the Fermi velocity $v_f=3.1 \cdot 10^7$ cm/s and energy $E_f=18$ meV. It should be noted that the results of experimental studies of viscosity $\nu$~\cite{gusev2020} are in agreement with the theoretical estimations based on (\ref{2D_par}).

Table~\ref{Table} also includes typical values of the electron mobility $\mu$ in AlGaAs quantum wells established in the literature~\cite{heiblum1984}, while the relaxation time $\tau_{ip}$ is estimated from the relation with the mobility: $\tau_{ip}=\mu m/e$. (It should be noted here that certainly in modern AlGaAs heterostructures the electron mobility can be larger than typical values presented in Table~\ref{Table}, see, for example, Ref.~\cite{chung2021}).

\begin{table}[h]
\centering
\caption{\label{Table} Typical values of the mean time of the electron-electron collisions $\tau_{ee}$, the time of electron collisions with impurities and phonons $\tau_{ip}$, the kinematic viscosity $\nu$, and the electron mobility $\mu$, are given for the case of 2DESs based on GaAs/AlGaAs quantum well. In estimations of $\tau_{ee}$ and $\nu$ (\ref{2D_par}) it was assumed that the concentration has the value of $5\cdot 10^{11}$ cm$^{-2}$.}
\begin{tabular}{|c||c|c|c|c|}
\hline
 & $\mu$, cm$^2$/(V$\cdot$ s) & $\tau_{ip}$  &$\tau_{ee}$ & $\nu$, cm$^2$/s \\
\hline
    T=4 K & $10^6$ & 38 ps\,  & 100 ps \,   & 2.4$\cdot 10^{4}$\\
    T=77 K & $2\cdot 10^5$ & 7.5 ps\,  & 0.27 ps\, & 64 \, \\
\hline
\end{tabular}
\end{table}

Thus, the condition of $\tau_{ee} \ll \tau_{ip}$ is satisfied at temperatures near $T=77$ K, while it is violated for AlGaAs quantum wells at the liquid-helium temperature.

In the case of plasma waves with a finite frequency $\omega$, we have to consider another condition for the validity of the hydrodynamic approach, namely, $\omega \tau_{ee} \ll 1$ (see, for example, the conclusion section in Ref~\cite{svintsov2018}). It means that the average time of electron-electron collisions should be much less than the period of plasma oscillations. Hence, this condition imposes the following restriction on frequency: $f =\omega/(2\pi) \ll (2 \pi \tau_{ee})^{-1}\approx 0.6$ THz at $T=77$ K, where the value of $\tau_{ee}$ is taken from Table~\ref{Table}.

In addition, for the considered plasma waves, we neglected the 'collisional' damping due to the finite value of $\tau_{ip}$, usually described by the term $v/\tau_{ip}$ in the left-hand side of the first equation in (\ref{Eq:basic}). Indeed, this damping factor can be ignored provided that $\omega\tau_{ip} \gg 1$. Hence, given that $\tau_{ip}=7.5$ ps at $T=77$ K, we obtain the constraint condition: $f=\omega/(2 \pi) \gg (2 \pi \tau_{ip})^{-1}\approx 20$ GHz.  

Therefore, the indicated approach is applicable as long as the frequency $\omega$ satisfies the requirement of $\tau_{ip}^{-1} \ll \omega \ll \tau_{ee}^{-1}$. For a 2DES with the abovementioned parameters, at the liquid-nitrogen temperature, this limitation corresponds to the frequency range between $20$ GHz and $0.6$ THz.

Furthermore, in the right-hand side of the first equation in (\ref{Eq:basic}), we also neglect the electronic pressure and viscosity terms, appearing as $-\frac{3}{4}v_f^2\partial_x n(x,t)/n_0$ (see, for example, Ref.~\cite{Rudin1997}) and $\nu \Delta v$, respectively. Indeed, the electronic pressure becomes significant in plasma waves at large wave vectors (at a large charge density gradient) of the plasmons --- qualitatively, at $q\approx k_F$, where $k_F$ is the Fermi wave vector. However, the study assumes the 'classical' limit of $q \ll k_F$, that is why the pressure can be neglected.

Concerning the viscosity it should be mentioned that, on the whole, it can be neglected provided that the Reynolds number, $Re$, is much greater than unity. Considering the estimate of $Re\approx \lambda v/\nu$, where $\lambda$ is the wavelength of the plasma wave, let us evaluate $Re$ at the plasmon frequency $f= 200$ GHz. According to the dispersion equation (\ref{Eq:spectrum}), this frequency corresponds to the plasmon wavelength of $\lambda\approx 10\, \mu$m, which is a typical value for the usual parameters $n_0=5\cdot 10^{11}$ cm$^{-2}$, $m=0.066m_0$, the distance between the 2DES and the gate, $d=400$ nm, and the dielectric permittivity of GaAs $\kappa=12.8$. The typical value of viscosity $\nu$ at 77 K is $64$ cm$^2$/s (Table~\ref{Table}). Therefore, a velocity $v=5\cdot 10^5$ cm/s can result in $Re\approx \lambda v/\nu\approx 8$. Here, it should be noted that, in principle, the velocity in a 2DES  can reach $10^7$ cm/s, see, for instance, Ref.~\cite{narozhny2022,Shur1987}. Thus, in principle, the Reynolds number can exceed the value of one.

Since the viscosity term appears in the right-hand side of the first equation in (\ref{Eq:basic}) as $\nu \Delta v$, qualitatively, viscosity leads to the damping of plasma waves with the characteristic rate of $\nu q^2$, where $q$ is the plasmon wave vector. Hence, for the aforementioned parameters, $\nu q^2=\nu (2 \pi/\lambda)^2\approx 2.5\cdot 10^9$ s$^{-1}$, which is insignificant compared to the typical frequency of the given plasma wave of the order of 100 GHz. Therefore, viscosity has a negligible effect on the plasma waves in the system.

To summarize, the study relies on two main assumptions: (i) the classical limit of $q\ll k_F$ ($\omega \ll E_F/\hbar$) and (ii) the hydrodynamic limit, requiring $\tau_{ip}^{-1} \ll \omega \ll \tau_{ee}^{-1}$ and a large Reynolds number, $Re\approx \lambda v/\nu$. 

Before proceeding, we note that on the whole, the considered waves are the nonlinear generalization of linear 2D plasmons in the hydrodynamic regime $\omega \tau_{ee} \ll 1$.

\section{Estimations of plasma wave parameters for realistic 2DES}\label{EST}

Thus, the possible experimental study of the considered plasma waves can be set up using GaAs/AlGaAs quantum wells (since in such structures, the electron relaxation time due to collisions with impurities and phonons is large) at or near the liquid-nitrogen temperature of $77$ K for the plasma frequencies on the order of 50 -- 300 GHz (higher frequencies may be used at a shorter relaxation time due to electron-electron collisions). The experimental investigation probably should be started with the measurements of the waves in the ungated 2DESs or systems with large $d$. In such systems, the plasmon wavelength is larger than in gated ones. For example, for the parameters mentioned above (the concentration $n_0=5\cdot 10^{11}$ cm$^{-2}$, the effective mass $m=0.066m_0$, and the dielectric permittivity of GaAs $\kappa=12.8$), at the frequency $f=200$ GHz, the wavelength becomes on the order of $\lambda=35\, \mu$m. In that case, the Reynolds number $Re\approx \lambda v/\nu$ equals 5.5 (see Table~\ref{Table} for typical $\nu=64$ cm$^2$/s) for the adequate velocity of electron flow, $v=10^5$ cm/s, which leads to the current density $j=e n_0 v = 0.8$ A/m. However, likely the latter condition is not very strict. For instance, a fivefold smaller velocity $v=2\cdot 10^4$ cm/s and the current density $j=e n_0 v = 0.16$ A/m will result in the Reynolds number on the order of 1. It means that qualitatively, the obtained by the analytical (hydrodynamic) approach results will still be applicable, although in this case the influence of viscosity may certainly require more detailed consideration.

Concerning the separation distance, $d$, note that it can vary over a fairly wide range, roughly from a relatively small distance of 0.5 $\mu$m \cite{Muravev2012} to a large distance of 500 $\mu$m \cite{Andreev2021}. Let us qualitatively characterize below solitons in the both cases.

For a large distance, $d=300$ $\mu$m, given typical frequencies of plasma waves within 50--300 GHz, the plasmon wavelength, $\lambda=2\pi/q$, ranges from 600 $\mu$m down to 17 $\mu$m, resulting in the values of $qd$ from 3 to 113. Since in this case $qd>1.61$, bright solitons can appear in the 2D system. An example of the bright soliton profile is shown in Fig.~\ref{fig:solitons}b. Considering the above-mentioned parameters at $q=5/d$ (given the rest of the parameters as indicated in the caption to Fig.~\ref{fig:solitons}b), Eqs.~(14) and (15) yield the soliton localization length of approximately $4d$ and the amplitude of $0.4n_0$. Here it should be noted, that the excitation and detection of solitons likely require the use of ohmic or capacitive side contacts to 2DES, so the distance between the contacts (together with the lateral size of the real 2D system) should certainly exceed the localization length (in our case $4d=1.2$ mm). 2D systems of such lateral size are indeed practically feasible. In Ref.~\cite{Gusikhin2018}, for instance, plasmons were experimentally studied in 2D disks (based on GaAs/AlGaAs quantum wells) as large as 1 cm in diameter.

As for the solitons in a gated 2D system, with small separation between the 2DES and the gates, $d=1$ $\mu$m, for the same frequency span of 50 to 300 GHz, the plasmon wavelength ranges from 61 $\mu$m down to 9.5 $\mu$m. In this case, the values of the parameter $qd$ fall between 0.1 and 0.66, resulting in the dark solitons that should arise in the system at $qd<1.61$. A typical profile of these solitons is shown in Fig.~\ref{fig:solitons}a. Thus, considering, for example, a soliton with $q=0.6/d$ (given the rest of the parameters as specified in the caption to Fig.~\ref{fig:solitons}a), Eqs. (12) and (13) lead to the soliton length of approximately $20d=20$ $\mu$m and the amplitude of $0.1n_0$. A soliton with such parameters seems to be quite realizable experimentally.

Thus, the above estimates show that the considered plasma waves can exist in realistic 2DESs based on GaAs/AlGaAs quantum wells.

As for the types of 2D systems beyond those involving quantum wells, graphene-based systems seem the most promising. For one thing, it is well-known that the hydrodynamic regime for the electron motion can be achieved in graphene~\cite{Muller2009,Bandurin2016,Narozhny2017}. Besides, the graphene-based structures with two gates are technically feasible and are frequently studied, for example, in the case of bilayer graphene (see, for instance, Ref.~\cite{oostinga2008}, note that in this reference the doped Si substrate acts as one of the gates). These reasons make graphene-based systems a good candidate for the possible investigation of the plasma waves considered in the paper.

\section{Discussion and conclusion}

Concerning the NLSE derivation by the multiple scales method mention once again that the corrections $n_2,n_3,...$ in (\ref{variables}) should be finite. However, the expression for $n_2$ has the denominator $G_0(q)-G_0(2q)$, which vanishes in the limiting case $qd \to 0$, see Eqs. (\ref{Eq:n2}) and (\ref{Coef_D}). The reason is that the plasmon dispersion (\ref{Eq:spectrum}) becomes linear for $qd \to 0$. In this case, the condition $\omega(2q)=2\omega(q)$ is satisfied, corresponding to the resonant second harmonic excitation, which is certainly beyond the application range of the NLSE. Qualitatively, the more nonlinear the dispersion is, the better the NLSE describes the behavior of the waves.

At the end it should be noted that the separation distances between the 2DES and the top and bottom gates do not necessarily have to be identical. The results reported in the paper are qualitatively valid regardless of the slight variation in the distances separating the 2DES and the top and bottom gates. However, in that case, the Poisson equation in (\ref{Eq:basic}) becomes rather intricate and cumbersome, and its analysis is beyond the scope of the paper. If necessary, this equation can be examined elsewhere, see, for example, Ref.~\cite{Govorov1999}.

To conclude, we have examined the nonlinear dynamics of quasi-monochromatic plasma waves in a 2DES placed between the two ideal metal gates. Employing the method of multiple scales, we derived the nonlinear Schr\"odinger equation (NLSE) governing the evolution of the slow-varying envelope of the waves. We found the NLSE solutions to be distinctly different depending on the parameter $qd$, where $q$ is the carrier wave vector and $d$ is the distance between the 2DES and the gates. Namely, when $qd<1.61$, the plane waves are stable and the NLSE solutions are dark solitons emerging against the plane-wave background. On the other hand, for $qd>1.61$, the plane waves are unstable due to the long-wavelength perturbations and the NLSE has bright-soliton solutions. The discovered phenomenon is remarkably similar to the behavior of water-surface waves~\cite{hasimoto1972}, where the conditions of shallow and deep water are distinguished based on whether the parameter $qh$ (with $h$ being the water depth) is below or above the critical value of $1.363$. We provide explicit NLSE solutions for the cases of dark and bright solitons as well as for the Peregrine breather, a peculiar space-time localized wave with a single maximum. We believe that the waves reported in the paper can be excited and detected in a real 2DESs based on the GaAs/AlGaAs quantum wells with side contacts at liquid nitrogen temperatures and frequencies of the order of 50 -- 300 GHz.

\begin{acknowledgments}
The author is grateful to Igor Zagorodnev and Vladimir Volkov for valuable discussions. The author also kindly thanks Viktor Ruban and Igor Burmistrov for helpful comments regarding nonlinear Schr\"odinger equation. The work was financially supported by the Russian Science Foundation (Project No. 21-72-00114).
\end{acknowledgments}

\appendix
\section{Derivation of the Green's function $G(q)$}\label{App:Green}
Given a 2DES placed between two ideal metal gates, as illustrated in the inset of Fig.~\ref{fig:nonlinear}, consider the Poisson equation relating the electron density in the 2DES, $\rho(x,y,t)$, and the electric potential, $\varphi(x,y,z,t)$:
\begin{equation} 
\label{AppPeq}
    -\Delta \varphi(x,y,z,t)=\frac{4\pi}{\kappa}\rho(x,y,t)\delta(z),
\end{equation}
where $x$ and $y$ are coordinates in 2DES plane, $z$-axis is perpendicular to 2DES plane as shown in the inset of Fig.~\ref{fig:nonlinear}, $t$ designates time, and $\kappa$ is the dielectric permittivity between the 2DES and the gates. Here 2DES is assumed to be $\delta$-thin. Taking the Fourier transform of this equation with respect to $x$ and $t$ yields $\varphi$ and $\rho$ as functions of the wave vector $q$ and the frequency $\omega$. We note that in this case, the solutions are homogeneous along the $y$-axis, i.e. the wave vector is directed along the $x$-axis. Therefore, Eq.~(\ref{AppPeq}) becomes:
\begin{equation}
\label{AppPeq2}
    (-\partial_z^2+q^2)\varphi(q,z,\omega)=\frac{4\pi}{\kappa}\rho(q,\omega)\delta(z).
\end{equation}
This differential equation is solved for the following boundary conditions along the $z$-coordinate: (i) the plasmon potential $\varphi(q,z,\omega)$ has to vanish at the surfaces of the ideal metal gates, i.e. $\varphi(q,z=\pm d,\omega)=0$, (ii) $\varphi(q,z,\omega$) has to be continuous at the location of the 2DES plane ($z=0$), and (iii) the partial derivative $\partial_z\varphi(q,z,\omega)$ at $z=0$ has to satisfy the equation below, obtained by integrating Eq.~(\ref{AppPeq2}) across the point $z=0$: 
\begin{equation}
\label{AppBC}
    \partial_z\varphi(q,z,\omega)|^{z=+0}_{z=-0}=-\frac{4 \pi}{\kappa} \rho(q,\omega).
\end{equation}
From the boundary conditions (i) and (ii), we find the solution for the potential $\varphi(z)$ of the form: 
\begin{equation}
    \varphi(q,z,\omega)=\varphi(q,z=0,\omega)\frac{\sinh(q(d-|z|))}{\sinh(qd)}.
\end{equation}
Finally, using the third boundary condition in (\ref{AppBC}) for the derivative $\partial_z \varphi(q,z,\omega)$ at $z=0$, we arrive at:
\begin{equation}
\label{AppPeq3}
     \varphi(q,z=0,\omega)=\frac{2\pi\rho(q,\omega)}{q \kappa}\tanh(qd).
\end{equation}
After applying the inverse Fourier transformation, the above equation becomes the 3rd equation in (\ref{Eq:basic}), where the plasmon charge density is related to the concentration as $\rho(x,t)=-en(x,t)$, where $e$ is the elementary charge ($e>0$). We also note that function $G(q)$ in Eq.~(\ref{Eq:Gf}) relates $\varphi(q,z=0,\omega)$ and $\rho(q,\omega)/\kappa$, therefore it is given as $G(q)=2\pi \tanh(qd)/q$. 

To determine the spectrum of linear plasmons, the first two expressions in~(\ref{Eq:basic}) need to be linearized. Hence, their Fourier transformation leads to:
\begin{equation}
\begin{split}
     -i\omega v(q,\omega) =iq\frac{e}{m}\varphi(q,z=0,\omega), \\
     -i\omega \rho(q,\omega) -iq en_0 v(q,\omega)=0,
\end{split}
\end{equation}
where $v(q,\omega)$ is the Fourier transform of the velocity of the electron flow, $m$ is the effective mass of electrons, and $n_0$ is the equilibrium concentration of electrons in 2DES. Using these equations along with~(\ref{AppPeq3}), we find the dispersion law given by Eq.~(\ref{Eq:spectrum}).

Here it is worth stressing the fundamental difference between plasmons in 3D and 2D systems. In the former case the plasmon frequency, $\omega_{3D}$, in the long wavelength limit, is independent of the value of the wave vector $q$: $\omega_{3D}=\sqrt{4 \pi e^2 n_{3D}/m}$, where $n_{3D}$ is the 3D concentration of electrons. In contrast, the frequency of plasmons in a 2D system (\ref{Eq:spectrum}) strongly depends on the wave vector $q$. Such a difference between the dispersions can be qualitatively explained as follows. Roughly speaking, 3D plasmons can be viewed as oscillations of positively and negatively charged planes. The electric field between the planes, corresponding to the restoring force, is independent of the distance between the planes. For this reason, the frequency of 3D plasmons is unrelated to the wave vector. As opposed to the charged planes, 2D plasmons (in a 2DES without the gates) can be treated as oscillations of charged 'wires'. In this case, the electric field between the wires and the restoring force behave as $1/r$, where $r$ is the distance between the wires. Hence, strong dependence of the frequency on the wave vector appears in the dispersion law of 2D plasmons~(\ref{Eq:spectrum}).

In the case of a gated 2DES, at $qd \ll 1$ (where $d$ is the distance between the 2DES and the gates), the Coulomb interaction between electrons in the 2DES is weakened due to the appearance of so-called image charges in the gates. As a result, the square-root plasmon spectrum in ungated 2DES is ‘softened’, exhibiting linear dependence: $\omega^{gated}(q)=v_p q$, with the velocity $v_p=\sqrt{2\pi e^2 n_0 d/(m\kappa)}$.

\section{Explicit expressions for $M_{0,1,2}$ and $L_{0,1,2}$}\label{App:ML}
Let us write down the Fourier-transformed Poisson equation~(\ref{AppPeq3}) in the following way:
\begin{equation}
\label{Eq:P2}
    \varphi(q,t) \cosh (qd) =-\frac{2\pi e d}{\kappa} n(q,t) \frac{\sinh{qd}}{qd}.
\end{equation}
Then, we can expand $\cosh (qd)$ and $\sinh (qd)/qd$ into Taylor series. Note that these series have an infinite radius of convergence with respect to $qd$ and include only the even powers of $qd$. Hence, after the inverse Fourier transformation, as $q^2$-terms become $-\partial^2_x$ operators, we obtain:
\begin{equation}
\label{Eq:Psplit}
    \widehat{M}\varphi(x,t) =-\frac{2\pi e d}{\kappa} \widehat{L} n(x,t),
\end{equation}
where the differential operators, $\widehat{M}$ and $\widehat{L}$, are given as:
\begin{equation}
\label{Dif_ML}
    \widehat{M}= \sum\limits^{\infty}_{j=0} \frac{(-1)^j d^{2j} \partial_x^{2j}}{(2j)!},\,\,
    \widehat{L}= \sum\limits^{\infty}_{j=0} \frac{(-1)^j d^{2j} \partial_x^{2j}}{(2j+1)!}.
\end{equation}

Next, based on Eqs.~(\ref{Eq:Dif}) and (\ref{Dif_ML}), we can determine the expansions for $\widehat{M}$ and $\widehat{L}$ in the $\varepsilon$-series as follows: 
\begin{equation}
\label{Dif_ops}
\begin{split}
 &\widehat{M} = \widehat{M}_0+\varepsilon \widehat{M}_1+\varepsilon^2 \widehat{M}_2 +...,\\
 &\widehat{L} = \widehat{L}_0+\varepsilon \widehat{L}_1+\varepsilon^2 \widehat{L}_2 +...
\end{split}
\end{equation}

To determine the operators $\widehat{M}_{0,1,...}$ and $\widehat{L}_{0,1,...}$ in the expansions~(\ref{Dif_ops}), it is convenient to use the Fourier transformations. Hence, in $L(q)=\sinh(qd)/qd$ and $M(q)=\cosh(qd)$, the argument $q$ can be expressed by the series $q_0+\varepsilon q_1+\varepsilon^2 q_2+...$, corresponding to the substitution in (\ref{Eq:Dif}). Then, we can expand these functions into the Taylor series and extract the coefficients of the different powers of $\varepsilon$, thereby finding $M_{0,1,..}(q_k)$ and $L_{0,1,..}(q_k)$ (with $k=0,1,...$) --- the Fourier transforms of the desired operators in~(\ref{Dif_ops}). As a result, for example, $M_{0,1,2}(q_k)$ and $L_{0,1,2}(q_k)$ can be written explicitly as follows:
\begin{equation}
\label{explicit_ML}
\begin{split}
    & M_0 = \cosh(q_0 d), \\ 
    & M_1 = q_1 d \sinh(q_0 d), \\
    & M_2= q_2 d \sinh(q_0 d) +\frac{q_1^2 d^2}{2}\cosh(q_0 d),\\
    & L_0 = \sinh(q_0 d)/q_0 d, \\
    & L_1= \frac{q_1}{q_0}\cosh(q_0 d)-\frac{q_1}{q_0}\sinh(q_0 d),\\
    & L_2  = \left(\frac{q_1^2 d}{2q_0} -\frac{q_2}{q_0^2 d} +\frac{q_1^2}{q_0^3 d} \right)\sinh(q_0 d) +\left( \frac{q_2}{q_0} \right. \\ 
    &  \quad \quad \left.  -\frac{q_1^2}{q_0^2}\right)\cosh(q_0 d) .\\
\end{split}
\end{equation}

Finally, to obtain the operators $\widehat{M}_{0,1,2}$ and $\widehat{L}_{0,1,2}$, we can expand the expressions in~(\ref{explicit_ML}) into the Taylor series and make the substitution $q_k \to -i\partial_{x_k}$, where $k=0,1,2,...$.

Note that $\widehat{M}_0$ and $\widehat{L}_0$ above have the simple form of the operators in (\ref{Dif_ML}), with $\partial_x$ replaced by $\partial_{x_0}$.

\section{Explicit forms of the key equations}\label{App:EE}
Retaining in (\ref{Eq:basic}) and (\ref{Eq:Psplit}) only terms proportional to $\varepsilon$, we arrive at:
\begin{equation}
\label{Eq:e1}
\left\{ 
\begin{split}
    & \partial_{t_0} v_1 -\partial_{x_0} \overline{\varphi}_1 =0  \\
    & \partial_{t_0} n_1 +n_0\partial_{x_0} v_1 =0 \\
    &\widehat{M}_0 \overline{\varphi}_1+ \frac{v_p^2}{n_0} \widehat{L}_0 n_1 =0   
    \end{split}
\right. 
\end{equation}
where $v_p=\sqrt{2 \pi e^2 n_0 d /(m\kappa)}$ is the velocity of the linear plasma waves (\ref{Eq:spectrum}) in the long-wavelength limit $qd \ll 1$ and $\overline{\varphi} = e\varphi/m$ is introduced to simplify the form of the equations.
Here, $v_1$ and $\overline{\varphi}_1$ in (\ref{Eq:e1}) can be eliminated by expressing them in terms of $n_1$, resulting in:
\begin{equation}
\label{e1_Eq}
    \left(-\widehat{M}_0 \partial_{t_0}^2 +v_p^2\partial_{x_0}^2 \widehat{L}_0\right)n_1=0,
\end{equation}
which after the substitution~(\ref{defin_n}) leads to Eq.~(\ref{eps1}). Also, it is straightforward to determine $v_1$ and $\overline{\varphi}_1$ from Eq.~(\ref{Eq:e1}) as $v_1=\omega n_1/(q n_0)$ and $\overline{\varphi}_1=-\omega^2 n_1/(q^2 n_0)$.

Now, let us proceed to the equations that follow from the vanishing of the coefficient of $\varepsilon^2$ in Eqs.~(\ref{Eq:basic}) and (\ref{Eq:Psplit}). Thus, after the substitution of the expressions from (\ref{variables}), (\ref{Dif_ML}) and (\ref{Dif_ops}), we obtain: 
\begin{equation}
\label{Eq:e2}
\left\{ 
\begin{split}
    & \partial_{t_0} v_2 +\partial_{t_1} v_1  -\partial_{x_0} \overline{\varphi}_2 -\partial_{x_1} \overline{\varphi}_1 + \frac{1}{2}\partial_{x_0} v_1^2=0  \\
    & \partial_{t_0} n_2 +\partial_{t_1} n_1 +n_0\partial_{x_0} v_2 +n_0\partial_{x_1} v_1 + \partial_{x_0} (n_1 v_1)=0 \\
    &\widehat{M}_0 \overline{\varphi}_2+\widehat{M}_1 \overline{\varphi}_1+ \frac{v_p^2}{n_0} \left( \widehat{L}_0 n_2+ \widehat{L}_1 n_1 \right)=0. 
    \end{split}
\right.
\end{equation}
Similar to~(\ref{e1_Eq}), the elimination of $v_2$ and $\overline{\varphi}_2$ leads to the following equation for $n_2$:
\begin{equation}
\label{Eqn2}
\begin{split}
        \left(-\widehat{M}_0 \partial_{t_0}^2 +v_p^2\partial_{x_0}^2 \widehat{L}_0\right) n_2= - n_0 \widehat{M}_0\partial_{x_0}(\partial_{t_1}v_1-\partial_{x_1}\varphi_1)  \\
        -\frac{n_0}{2} \widehat{M}_0\partial_{x_0}^2 v_1^2 +\widehat{M}_0\partial_{t_0}\left( \partial_{t_1}n_1+n_0 \partial_{x_1} v_1 +\partial_{x_0}(n_1 v_1)\right) \\
        -n_0 \widehat{M}_1 \partial_{x_0}^2 \varphi_1 -   v_p^2 \widehat{L}_1 \partial_{x_0}^2 n_1.
\end{split}
\end{equation}
The main point of the multiple scales method is that all the corrections $n_2, n_3,...$ should be finite, without unbounded growth at $t_0 \to \pm \infty$ or $x_0 \to \pm \infty$. That is why no 'resonant' (secular) terms $\propto \exp(i\theta)$, where $\theta=q x_0-\omega t_0$, are expected in the right-hand side of the equation for corrections. Namely, setting to zero the coefficient of the term $\propto \exp(i\theta)$ in the right-hand side of Eq.~(\ref{Eqn2}), one can obtain Eq.~(\ref{envelope_vel}).

Then, we consider $n_2$ of the form:
\begin{equation}
\label{Eq:n2}
    n_2= C + D \exp(2i\theta) +D^* \exp(-2i\theta),
\end{equation}
where coefficients $C$ and $D$ are the functions of only the slow variables $x_1$, $x_2$, $t_1$, $t_2$ and so on, and $C$ has a real value. Consequently, $n_2$ has the 'constant' and the second harmonic terms. It is also assumed that $n_2$ has no eigen contribution $\propto \exp(i\theta)$, more details on the issue are given below, after Eqs.~(\ref{App:v_phi}) and (\ref{App:BB}). The $D$ coefficient can be found by introducing the relation in (\ref{Eq:n2}) into Eq.~(\ref{Eqn2}) as follows:
\begin{equation}
\label{Coef_D}
    D=\frac{3A^2}{2n_0}\frac{G_0(q)}{G_0(q)-G_0(2q)}, \, \text{where} \,\,G_0(q)=\frac{\tanh qd}{qd}.
\end{equation}
Although the coefficient $C$ cannot be determined from the equations for $\varepsilon^2$, it will be defined based on the equations for $\varepsilon^3$ as discussed below.

Along with the desired type of $n_2$ in (\ref{Eq:n2}), we consider $v_2$ and $\overline{\varphi_2}$ of the following forms:
\begin{equation}
\label{App:v_phi}
\begin{split}
     v_2= C_v + B_v \exp(i\theta) + D_v \exp(2i\theta) + c.c.,\\
     \overline{\varphi}_2= C_{\varphi} + B_{\varphi} \exp(i\theta) + D_{\varphi} \exp(2i\theta) +c.c.,
\end{split}
\end{equation}
where $C_v$ as well as $C_{\varphi}$ have real values and all the coefficients depend only on the slow variables, $x_1$, $x_2$, $t_1$, $t_2$, and so on. It should be noted that $v_2$ and $\overline{\varphi}_2$ above may have a contribution $\propto \exp(i\theta)$.

By introducing the expressions in (\ref{App:v_phi}) into Eqs.~(\ref{Eq:e2}) and using the definition of the coefficient $D$ in Eq.~(\ref{Coef_D}), we arrive at:
\begin{equation}
    \begin{split}
        D_v= \frac{\omega A^2}{q n_0^2}\frac{G_0(q)/2 +G_0(2q)}{G_0(q)-G_0(2q)}, \\
        D_{\varphi}= -\frac{\omega^2 A^2}{q^2 n_0^2}\frac{3G_0(2q)/2}{G_0(q)-G_0(2q)},
    \end{split}
\end{equation}
and 
\begin{equation}
\label{App:BB}
    \begin{split}
        B_v= \frac{i}{q n_0}\left(\partial_{t_1} A +\frac{\omega}{q} \partial_{x_1} A \right), \\
        B_{\varphi}= -\frac{2i\omega}{q^2 n_0}\left(\partial_{t_1} A +\frac{\omega}{q} \partial_{x_1} A \right).
    \end{split}
\end{equation}
Now, adding an 'eigen' term $B\exp(i\theta)$ (with an arbitrary value of amplitude $B$) to $n_2$~(\ref{Eq:n2}) leads to extra contributions to $B_v$ and $B_{\varphi}$ in the forms of $\omega B/(qn_0)$ and $-\omega^2 B/(q^2 n_0)$, respectively. We also note that from the equation for $\varepsilon^3$ below, an expression analogous to Eq.~(\ref{envelope_vel}) can be obtained for the coefficient $B$.

The equations corresponding to the vanishing of the coefficients at $\varepsilon^3$ can be written as:
\begin{widetext}
\begin{equation}
\label{App:e3}
\left\{ 
\begin{split}
    & \partial_{t_0} v_3 +\partial_{t_1} v_2 +\partial_{t_2} v_1 -\partial_{x_0} \overline{\varphi}_3 -\partial_{x_1} \overline{\varphi}_2 -\partial_{x_2} \overline{\varphi}_1+ \frac{1}{2}\partial_{x_1} v_1^2 +\partial_{x_0}(v_1 v_2)=0  \\
    & \partial_{t_0} n_3 +\partial_{t_1} n_2 +\partial_{t_2} n_1 +n_0\partial_{x_0} v_3 +n_0\partial_{x_1} v_2 +n_0\partial_{x_2} v_1+ \partial_{x_1} (n_1 v_1) +\partial_{x_0}(n_1 v_2+  v_1 n_2)=0 \\
    &\widehat{M}_0 \overline{\varphi}_3+\widehat{M}_1 \overline{\varphi}_2+ \widehat{M}_2 \overline{\varphi}_1+\frac{v_p^2}{n_0} \left( \widehat{L}_0 n_3+ \widehat{L}_1 n_2 +\widehat{L}_2 n_1\right)=0   
    \end{split}
\right.
\end{equation}
\end{widetext}

\onecolumngrid

Extracting the slow terms, that do not depend on $x_0$ and $t_0$, in the first two equations in~(\ref{App:e3}) and in the third equation in~(\ref{Eq:e2}) yields the following expressions involving the coefficients $C$, $C_v$, and $C_{\varphi}$:
\begin{equation}
\label{App:const}
\left\{ 
\begin{split}
    & \partial_{t_1} C_v -\partial_{x_1} C_{\varphi} +\frac{\omega^2}{2 q^2 n_0^2} \partial_{x_1}|A|^2=0  \\
    & \partial_{t_1} C +n_0\partial_{x_1} C_v +\frac{2\omega}{qn_0} \partial_{x_1}|A|^2 =0 \\
    & M_0(q=0) C+ \frac{v_p^2}{n_0} L_0(q=0) C_{\varphi} =0, 
\end{split}
\right.
\end{equation}
where $M_0(q=0)= L_0(q=0)=1$. Then, using Eqs.~(\ref{App:const}) together with the relation $\partial_{x_1} |A|^2=v_{gr}^2 \partial_{t_1} |A|^2$ that follows from Eq.~(\ref{envelope_vel}), we can define the coefficients $C$, $C_v$, and $C_{\varphi}$ as follows:
\begin{equation}
\begin{split}
\label{App:const_fin}
    & C = -\frac{\omega}{q n_0}\left(2v_{gr}+\frac{\omega}{q} \right) \frac{|A|^2}{v_p^2 - v_{gr}^2}, \\
    & C_v = -\frac{\omega }{q n_0^2} \left(2v_p^2+v_{gr}\frac{\omega}{q} \right) \frac{|A|^2}{v_p^2 - v_{gr}^2},
    \end{split}
\end{equation}
and $C_{\varphi}=-v_p^2 C/n_0$.

Next, by eliminating $v_3$ and $\overline{\varphi}_3$ in Eqs.~(\ref{App:e3}) we arrive at the following equation for $n_3$:
\begin{equation}
\label{App:Eqn3}
\begin{split}
        \left(\widehat{M}_0 \partial_{t_0}^2 -v_p^2\partial_{x_0}^2 \widehat{L}_0\right) n_3=
        -\widehat{M}_0\partial_{t_0} \left(\partial_{t_1} n_2+\partial_{t_2} n_1+ n_0 \partial_{x_1}v_2 + n_0 \partial_{x_2} v_1 +\partial_{x_1}(n_1 v_1)+\partial_{x_0}(n_1 v_2+ v_1 n_2)  \right) \\
        +n_0\widehat{M}_0\partial_{x_0}\left(\partial_{t_1}v_2+\partial_{t_2}v_1 -\partial_{x_1}\overline{\varphi}_2 -\partial_{x_2} \overline{\varphi}_1 +\frac{1}{2}\partial_{x_1}v_1^2 +\partial_{x_0}(v_1 v_2) \right) +n_0 \widehat{M}_1 \partial_{x_0}^2 \varphi_2 +n_0 \widehat{M}_2 \partial_{x_0}^2 \varphi_1 \\
        + v_p^2 \partial_{x_0}^2 \left(\widehat{L}_1  n_2 +\widehat{L}_2  n_1 \right).
\end{split}
\end{equation}
Finally, setting to zero the coefficient of the secular term $\propto \exp (iqx_0 - i\omega t_0)$ in the right-hand side of the above equation results in the desired nonlinear Schr\"odinger equation~(\ref{NLSE}).

\twocolumngrid

\end{document}